# Nuclear Attenuation of Fast Hadrons Produced in Charged-Current $\nu$ and $\bar{\nu}$ Interactions in Neon

## BEBC WA21/WA59 Collaborations

W. Burkot[1,a], T. Coghen[1,b], J. Czyżewski[1,c], M. Aderholz[2], J. Guy[3], G.T. Jones[4], U.F. Katz[2,†], P. Marage[5], D.R.O. Morrison[6], G. Myatt[7], N. Schmitz[2], W. Venus[3], W. Wittek[2]

[1] The M. Miesowicz Inter-Institute Centre for High Energy Physics, PL-30055 Cracow, Poland
   a) Academy of Mining and Metallurgy
   b) Institute of Nuclear Physics
   c) Jagellonian University
[2] Max-Planck-Institut für Physik, D-80805 München, Germany
[3] Rutherford Appleton Laboratory, Chilton, Didcot OX11 0QX, United Kingdom
[4] Department of Physics, University of Birmingham, Birmingham B15 2TT, United Kingdom
[5] Inter-University Institute for High Energies, ULB-VUB, B-1050 Brussels, Belgium
[6] CERN, CH-1211 Geneva 23, Switzerland
[7] Department of Nuclear Physics, University of Oxford, Oxford OX1 3RH, United Kingdom
† Now at Physikalisches Institut, Universität Bonn, D-53115 Bonn, Germany

Revised Version
November 1995

**Abstract**

The production of hadrons in charged-current (anti)neutrino interactions is studied with the bubble chamber BEBC exposed to the CERN (anti)neutrino wide-band beam. Fast-hadron production in a neon target is found to be attenuated as compared to that in a hydrogen target. This feature is discussed within theoretical models based on the idea of a hadron formation length. The experimental results favour the 'constituent' over the 'yo-yo' length concept, and suggest a quark cross section in the order of 3 mb.

# 1  Introduction

The study of hadron production in deep-inelastic lepton-nucleus collisions provides information on the space-time development of the hadronization process, in particular on the so-called formation time or corresponding formation length. Indeed, secondary hadrons instantaneously produced at the interaction point would undergo strong rescattering (reinteractions) when traversing the dense nuclear matter on their way out of the nucleus. This leads to a dissipation of their energy. However, the hadrons may not be created at the interaction point, the produced object (e.g. the struck quark) travelling some distance in nuclear matter before fragmenting into hadrons. The amount of rescattering of these hadrons inside the nucleus and the dissipation of their energy would thus be reduced in the case of long 'formation lengths'. It follows from the uncertainty principle and from time dilatation that the formation time is proportional to the energy $E_h$ of the hadron [1]. This simple idea has been discussed in [2], and reworked in [3]–[5] in the context of the Lund string model of hadronization [6, 7]. Although the asymptotic hadronic state becomes defined after the time defined by $E_h$, the composite nature of hadrons causes an ambiguity in the definition of the formation time, since different constituents of the hadron may appear at different times. If they are able to interact in nuclear matter their formation time may be relevant for the energy dissipation, even if the formation of the hadrons is completed outside the nucleus.

Two extreme cases are given by the models (see Sect. 2) in which the relevant formation length $l_f$ is assumed to be given by the constituent length or the yo-yo length [3]. The constituent length $l_c$ corresponds to the time after which the first constituent of the hadron has been formed; the yo-yo length $l_y$ corresponds to the time after which all hadron constituents meet to form an oscillating 'yo-yo' state.

The energy dissipation results in a 'depletion' (attenuation) of energetic (fast) hadrons for interactions with a nucleus as compared to interactions with an elementary target. This depletion (of the fastest hadrons) is not replenished by secondary hadrons as reinteractions mainly yield lower energy products. A depletion of this kind has been observed in $z$-distributions in electroproduction [8] and muoproduction [9, 10] experiments, and has also been reported by our collaborations in the rapidity distributions of negative hadrons in $\bar{\nu}$Ne interactions [11].

In this context, it is convenient to study charged-hadron production in terms of normalized $z$ distributions

$$D(z) = \frac{1}{N_{ev}} \cdot \frac{dN_{had}}{dz} , \qquad (1)$$

where $N_{ev}$ is the number events and $dN_{had}$ the number of charged hadrons with $z$-values in the bin $(z, z+dz)$. $z$ is defined as

$$z = \frac{E_h + p_h^L}{E_{hads} + P_{hads}} , \qquad (2)$$

where $E_h$, $p_h^L$ are the energy and the longitudinal momentum of the hadron with respect to the virtual-boson direction, and $E_{hads}$, $P_{hads}$ are the total energy and momentum of the hadronic system, all measured in the laboratory frame. $z$ is Lorentz-invariant with respect to boosts along the virtual-boson direction and is therefore not sensitive to the smearing due to a longitudinal movement of the target nucleon. Such longitudinal movements may be due to Fermi motion or may arise from nucleon-nucleon correlations [12]. The region $z \gtrsim 0.2$, where $z$ is close to the Feynman-$x$ ($x_F$) variable, contains in practice only particles going into the forward region ($x_F > 0$) of the virtual-boson-nucleon center-of-mass system (cms). (The forward region is defined by the 3-momentum vector of the virtual boson).

Using $z$ distributions which are normalized by the number of events ($N_{ev}$) has the advantage of being insensitive to initial-state-interaction effects like the EMC effect at high Bjorken $x$ or nuclear shadowing at small $x$, which manifest themselves in the ratios of the total cross sections (proportional to $N_{ev}$)[13]. Therefore, by comparing $D(z)$ for different targets one measures only those effects which are related to the interactions of the final-state hadronizing system in the nucleus.

In the following the ratio $R_A(z)$ will be used to discuss the attenuation of fast hadrons in a nucleus with mass number $A$. $R_A(z)$ is defined as the ratio of fragmentation functions

$$R_A(z) = D_A(z)/D_N(z) \qquad (3)$$

for a nuclear target $A$ and a nucleon $N$.

## 2  The theoretical model

In the framework of the Lund model, the authors of [5] have calculated $R_A(z)$ as a function of $z$, the leptonic energy transfer $\nu$ in the laboratory frame and the number of nucleons in the nucleus, for both $l_f = l_c$ and $l_f = l_y$. The dependence of $l_c$ and $l_y$ on $z$ is given by

$$\langle l_c \rangle = zL \left[ \frac{\ln(z^{-2})}{1 - z^2} - 1 \right] \qquad (4)$$

$$l_y = l_c + zL, \qquad (5)$$

where the string length $L$ is related to the total hadronic energy and momentum via the string tension $\kappa$

$$2L = \frac{E_{hads} + P_{hads}}{\kappa}. \qquad (6)$$

In the limit $z \to 1$, $\langle l_c \rangle$ tends to zero, since a $z = 1$ hadron contains the struck quark, which is already present at the very beginning of the fragmentation process. On the other hand, the yo-yo length $l_y$ becomes equal to the string length $L$ at $z = 1$. The differences in the predictions of the two models are thus largest at high $z$, i.e. for the fastest hadrons.

The model is a generalization for leptoproduction of the model developed in [3] for hadron-nucleus collisions. As was noted earlier (??? where ???), contrary to the hadron-nucleus case, when studying leptoproduction one does not need to take into account the initial-state interactions.

In that model the attenuation ratio $R_A(z)$ is calculated as the average over the nuclear volume and over the formation length distribution $g(z, l' - l)$ of the probability $W_0(l, l', \vec{b})$ that no inelastic final-state interaction occured:

$$R_A = \int d^2b \int_{-\infty}^{\infty} dl \, \rho(l, \vec{b}) \int_{l}^{\infty} dl' \, g(z, l' - l) \, W_0(l, l', \vec{b}). \qquad (7)$$

$\rho(l, \vec{b})$ is the nuclear density normalized to one and $(l, \vec{b})$ and $(l', \vec{b})$ denote the interaction point (longitudinal coordinate, impact parameter) of the virtual boson and the formation point of the hadron, respectively. $W_0$ is defined by the cross section for the inelastic interactions of the intermediate quark ($\sigma_q$) and the produced hadron ($\sigma_h$) on a nucleon:

$$W_0(l, l', \vec{b}) = \left(1 - \int_{l}^{l'} dl'' \, \sigma_q \, \rho(l'', \vec{b}) - \int_{l'}^{\infty} dl'' \, \sigma_h \, \rho(l'', \vec{b})\right)^{A-1}. \qquad (8)$$

In [3] the distributions of the constituent and the yo-yo formation lengths were calculated in the Lund model [6] using the string splitting function

$$f(z) = (1 + C)(1 - z)^C. \qquad (9)$$

The average values of these distributions are given in eqs. (4) and (5). The above splitting function yields a fragmentation function of the form $D(z) \propto f(z)/z$. In the present analysis $C$ was chosen as 1.8, which gives a good description of the shape of $D_p(z)$ both for neutrino and antineutrino interactions, in all $\nu$ ranges. The corresponding distributions (normalized to the experimental data for $H_2$) are drawn as dotted lines in Figs 1a and 1b.

The nuclear density was taken in the Woods-Saxon form with an average nuclear radius $r_a$ of $(1.19 \, A^{1/3} - 1.61 \, A^{-1/3})$ fm and a surface parameter $a$ of 0.54 fm [3]. For the string tension $\kappa$ a value of 0.8 GeV/fm was used [5].

For $\sigma_h$ the experimental value of the inelastic non-single-diffractive pion-nucleon cross section was taken [14]. Thus the slight energy dependence of this cross section present at low energies, and the variation of the cross section due to the heavier resonances or to the tails of the lighter ones are automatically taken into account. However, even at the lowest values of $z$ and $\nu$ considered in this analysis the secondary pion-nucleon interactions are almost outside the resonance region.

The quark cross section $\sigma_q$ is assumed either to be equal to zero or to be very small. As was argued e.g. in [15], the quark (which in the case of the leptoproduction is the current, undressed quark) cannot interact inelasticly in nuclear matter with a large cross section. The experimental data suggest $\sigma_q \sim 1$ mb at high energy [5]. A similar size of $\sigma_q$ was also used in [16] for high-energy leptoproduction data. However, in [16] it was argued that a slight rise of $\sigma_q$ can be present towards relatively small values of $Q^2$.

## 3  Data

The present paper summarizes the experimental analysis done in [17]. The analysis is based on $\nu$ and $\overline{\nu}$ charged-current (CC) interactions observed in the bubble chamber BEBC, filled with a 75 mole% Ne-$H_2$ mixture (WA59 experiment) or with $H_2$ (WA21 experiment) and exposed to the CERN SPS wide-band beam.

The (anti)neutrino beam in the WA59 experiment originated from 400 GeV protons impinging on a beryllium target. In order to ensure similar kinematic conditions in the two experiments only that part of the WA21 data was used in this analysis which was obtained with a 400 GeV proton beam.

For the calculation of the kinematic quantities in an event the laboratory energy $E_\nu$ of the incoming (anti)neutrino has to be known. In the present experiment $E_\nu$ cannot be directly determined since there is a substantial loss of energy due to undetected neutral particles. Various methods of correcting the visible energy based on transverse momentum balance were applied [18, 19] in order to obtain an estimate of $E_\nu$. $\nu$ was then determined as $E_\nu - E_\mu$, where $E_\mu$ is the laboratory energy of the muon. Identified protons with a momentum less than 1.2 GeV/c were not considered in the analysis, and the pion mass was assigned to all remaining observed charged hadrons. Neutrals from both the Ne and $H_2$ data sets were not used when estimating $E_\nu$, in order to minimize systematic differences between the two data samples due to the higher efficiency of neutral particle detection in neon.

The events were required

- to have an identified muon with the right charge ($-$ for $\nu$, $+$ for $\overline{\nu}$) and a momentum greater than 5 GeV/c,

- to contain in addition at least one charged hadron, and

- to have $Q^2 > 1$ GeV$^2$ and $W > 2$ GeV, where $Q^2$ is the negative square of the 4-momentum transfer from the lepton to the hadron vertex and $W$ is the invariant mass of the hadronic system.

In the WA59 experiment, candidates for coherent events [20] were removed by demanding the negative square of the 4-momentum transfer from the virtual photon to the neon nucleus to be greater than 0.1 GeV$^2$, for those events in which the total visible hadronic charge was equal to the virtual-boson charge.

After all selections the event samples consist of 8679 $\overline{\nu}$Ne, 6197 $\nu$Ne, 2429 $\overline{\nu}$p and 2648 $\nu$p events.

## 4 Corrections

Fig. 1 depicts the raw fragmentation functions $D(z)$ for charged hadrons from CC interactions of neutrinos and antineutrinos in neon (solid circles) and in hydrogen (open circles) for four $\nu$ intervals. Already in these raw data one observes an excess of hadrons at the smallest $z$, due to rescattering in the neon nucleus, in the distributions for neon as compared to those for hydrogen. The smaller excess in the neutrino samples at low $z$ can be attributed to constraints due to charge conservation, which result in a higher average backward multiplicity in $\nu$p as compared to $\overline{\nu}$p scattering [21]. At high $z$, on the other hand, no conclusions can be drawn from the raw data since Monte Carlo simulations show that due to strong smearing at high $z$ the observed $z$ distributions are distorted differently for H$_2$ and Ne. Therefore, the raw data have to be corrected for such experimental biases. The adopted unfolding procedure is described in the following.

The main source of systematic errors in the present analysis is the substantial loss of energy due to undetected neutral particles, affecting the estimation of the neutrino energy and thus also the calculation of most of the kinematic quantities. It was found that the results on fragmentation functions at high $z$ depend on the energy correction method and the binning in $z$ used.

A popular method to correct measurements for experimental resolution and systematic effects relies on estimating a ratio between 'true' and 'observed' distributions from a Monte Carlo simulation and on using this ratio to correct the measurements. It is known that this correction factor method is biased towards reproducing the initially assumed distribution. This is especially crucial if the smearing is large, which in the present analysis is the case in the interesting region of high $z$. Therefore an alternative unfolding method was applied, which is rather independent of the assumption about the true form of the initial distribution.

In general, the content $S_n$ of the $n$-th bin in the distribution of an observed quantity is related to the content $T_k$ of the $k$-th bin in the distribution of the true quantity by the detector response matrix $C$ (with elements $C_{nk}$):

$$S_n = \sum_k C_{nk} T_k \ . \qquad (10)$$

The true distribution is calculated by solving the so called inverse problem [22], simply in-

verting the matrix $C$:

$$T_k = \sum_n (C^{-1})_{kn} S_n \ . \tag{11}$$

The elements $C_{nk}$ of the response matrix were determined by a detailed Monte Carlo (MC) simulation of the experiment [23], taking into account the experimental resolution and all known sources of systematic effects (see below), in the following way.

Generated ('true') MC events were subjected to a simulation of the measuring and reconstruction process including all known systematics. The result from this procedure is referred to as 'observed' MC events. Seven bins were chosen for the generated and observed values of the leptonic energy transfer $\nu$ in the laboratory system (2 to 4 GeV, 4 to 8 GeV, 8 to 16 GeV, 16 to 32 GeV, 32 to 64 GeV, 64 to 128 GeV and $> 128$ GeV), and 10 bins of width 0.1 for the generated and observed values of $z$. The $70 \times 70$ matrix $C$ was then determined as

$$C_{nk} = \frac{N_{nk}}{\sum_i N_{ik}}, \tag{12}$$

where $N_{nk}$ is the number of hadrons with observed $(\nu, z)$ values in the $n$-th bin of the $\nu$-$z$ plane and true $(\nu, z)$ values in the $k$-th bin of the $\nu$-$z$ plane.

Hadrons produced in reinteractions in the nucleus mainly populate the $z < 0.2$ region and contribute only little to $z > 0.2$. Their number as well as their $z$ distribution depends on details of the reinteraction model used in the MC simulation. In a rough evaluation the contamination of the $z > 0.2$ hadron sample by hadrons from reinteractions is estimated to be less than (???) %. As no attempt was made to correct the $z < 0.2$ data points, which are affected strongest by the hadrons from reinteractions, only those tracks were included in $N_{nk}$ which did not originate from reinteractions.

Since in the middle of the $z$ range of interest the resolution in $z$ is about 0.2 (as can be seen from the example shown in Table 1), the final distributions in $z$, for a given $\nu$ interval, were determined by combining the results from neighbouring bins. This procedure was intended to reduce the scatter of points due to the statistical errors of the data.

It was found that with the chosen binning the results for the corrected $(\nu, z)$ distributions, obtained by applying the procedure (11) to the measured $(\nu, z)$ distributions, are stable against small variations of bin sizes. This observation and the fact that the corrected distributions do not exhibit unreasonable fluctuations are considered as a posteriori justification for using the simple procedure (11) without any regularisation.

A similar method of unfolding, however without rebinning, was used for correcting the number of events $N_{ev}$ in each $\nu$ interval.

The matrices $C$ were calculated for each projectile-target combination and for each energy correction method. Altogether four methods were used, namely the two methods described in [18] and [19] and two modifications of those methods. It was found that for a given theoretical distribution in $\nu$ and $z$ the 'observed' MC ratios $R_{\mathrm{Ne}}(z) = D_{\mathrm{Ne}}(z)/D_p(z)$ for the different energy correction methods exhibit significant differences between each other. Similar differences were observed in the experimental (uncorrected) data.

In the present case, the main sources of systematic differences between the Ne and $H_2$ data samples are the following:

- Differences in the result of the energy correction due to the Fermi motion and to reinteractions of hadrons in the neon nucleus affecting the transverse momentum balance of the event.

- Different precision on momentum measurement and different efficiencies of particle identification.

- Different scanning efficiency for low-multiplicity events.

- Nuclear effects such as coherent production [20] and shadowing [24]. Candidates for coherent production were removed by appropriate cuts (see above).

- Differences in the fragmentation functions due to the non-isoscalarity of the $H_2$ target used as a reference target.

All these effects have been simulated in the MC calculations and are thus taken into account in the corrections.

A global systematic error on $R_{Ne}$ was determined by varying details of the analysis, like the kinematical cuts and the energy estimation method. At $z = 0.8$ the systemaic error was estimated as $\leq 17\%$ and at $z = .05$ as $\leq 5\%$, the dominant contribution being due to uncertainties in the energy estimation.

Corrected results will be presented for the $\nu$ bins which contain sufficient statistics and which are free of edge effects. As has been pointed out before, the smearing matrix $C$ was constructed in a way which reduces the dependence on the particular reinteraction model in the MC simulation. This excludes a reliable calculation of corrections for the $z < 0.2$ data points. Therefore no corrected data are shown for $z < 0.2$.

## 5 Results

Figures 2 to 5 show the corrected (unfolded) results for $R_{\mathrm{Ne}}(z)$. The crosses on the error bars correspond to the unfolded $R_{\mathrm{Ne}}$ values from the two energy estimation methods [18] and [19], which in general gave the highest and lowest corrected $R_{\mathrm{Ne}}$ in the $z$ bin. Their difference illustrates the sensitivity of the unfolded results to the energy estimation method. The open circles represent the weighted average of the results from methods [18] and [19]. Only the results of those bins that contain at least 10 entries are shown. The vertical error bars represent the statistical errors.

A significant depletion of hadrons in the interactions with neon is seen in the whole region $0.2 < z < 1$ for $\nu \lesssim 16$ GeV (Figs. 2, 3). There is a similar trend for $\nu \gtrsim 16$ GeV, however, the statistical significance is low (Figs. 4, 5). Fig. 6 shows the $\nu$ dependence of the ratio $R_{\mathrm{Ne}}(z > 0.2) = \left( \int_{0.2}^{1} D_{\mathrm{Ne}}(z) dz \right) / \left( \int_{0.2}^{1} D_{\mathrm{p}}(z) dz \right)$ of the combined neutrino and antineutrino data.

The predictions of the models described in Section 2 are compared with the experimental data in Figs. 2 to 6. The solid and dashed-dotted lines are for the $l_c$ model, the dashed and dotted lines for the $l_y$ model. The dotted and dashed-dotted lines represent the predictions for $\sigma_q = 0$, the solid and dashed lines those for $\sigma_q = 3$ mb.

The depletion observed in the experimental data in the region $z > 0.2$ can be taken as evidence for either strong interactions of the fastest hadrons or for a non-vanishing quark cross section. It should be noted that a small contamination of the $z > 0.2$ hadron sample by hadrons from reinteractions in the Ne nucleus would tend to increase the measured $R_{Ne}$ and thus to diminish the observed attenuation.

From the data at high $\nu$ (figs. 4 to 6), where the predictions from the $l_c$ and $l_y$ model are similar, one can conclude that $\sigma_q$ is different from zero and in the order of 3 mb.

At low $\nu$ and high $z$, where the differences between the $l_c$ and the $l_y$ model are largest (figs. 2 and 3), the data seem to favor the $l_c$ over the $l_y$ model.

Since $R_{Ne}(z > 0.2)$ is dominated by intermediate-$z$ hadrons, the data in Fig. 6 test the two models in the region of intermediate $z$. Also in this case the $l_c$ model is preferred by the data at low $\nu$.

In the $l_y$ model, a change of $\sigma_q$ from 0 to 3 mb causes a decrease of $R_A$ by about 0.08, quite independent of $z$ and $\nu$. In the $l_c$ model, on the other hand, the same change in $\sigma_q$ leads to a reduction of $R_A$ only at lower $z$ while leaving the prediction for $z \to 1$ unchanged. The latter behaviour is due to the vanishing life time of the intermediate quark in the limit $z \to 1$ in the $l_c$ model .

The different sensitivity of the two models to a variation of $\sigma_q$ has implications on the $z$ and $\nu$ dependence of $R_A$. The $l_c$ model gives a very good description of both the $z$ and the $\nu$ dependence of the attenuation. On the contrary, the $l_y$ model predicts a too weak $z$ and $\nu$ dependence. This behaviour is not changed even if $\sigma_q$ in the $l_y$ model is taken large enough to account for the overall magnitude of the attenuation.

These observations confirm previous results of [3] and [5], concering the constituent formation point of the hadron and the small value of the quark cross section, obtained for the charged-lepton [9] and hadron-induced [25] reactions. The fact that significant inelastic interactions of the hadronizing system can start as early as at the constituent point could seem surprising. However, as was shown in [26], the elastic colour exchange interactions of quarks (endpoints of the string) in nuclear matter with a cross section of the order of 10 mb lead effectively to the picture where the strong absorption starts at the constituent point. The quantitative arguments for colour exchange to be a mechanism of the nuclear suppression in leptoproduction were given also earlier in [27]. In [15] and [28] it was shown why such interactions occuring between the lepton interaction point and the constituent point do not lead to any additional strong suppression and why the observed value of $\sigma_q$ is small.

## 6 Summary

In this analysis hadron production in (anti)neutrino charged-current interactions on neon nuclei is compared with that on a proton target. Smearing effects and systematic effects in the experimental data have been corrected for by using an unfolding procedure based on a detailed Monte Carlo simulation of the experiments.

The data exhibit a significant attenuation of fast charged hadrons over the whole $\nu$ range considered ($4 < \nu < 64$ GeV), with the strongest attenuation at low $\nu$ and high $z$. Two extreme theoretical models were considered: the $l_y$ model, where the hadron formation length is given by the yo-yo length, and the $l_c$ model, where the formation length is assumed to be equal to the constituent length. In both models the cross section of the intermediate quark was an additional parameter. The data favor the $l_c$ model and suggest a quark cross section in the order of 3 mb. The $\nu$ and $z$ dependence as well as the size of the attenuation are well described by the model.

*Acknowledgements.* We express our gratitude to the CERN staff for the operation of the SPS accelerator, neutrino beam, and BEBC with associated equipment, and to the scanning and measuring staff in our institutions for their dedicated work. One of us (T.C.) is grateful to the Werner-Heisenberg-Institut (MPI) for the kind hospitality and creative atmosphere during his stay in Munich where the idea of studying nuclear effects in the WA59 experiment originated. W.B. is grateful to the MPI for providing the opportunity to complete this paper in Munich.

**Table 1:** Some elements $C_{nk}$ of the response matrix (using method [19]) for events with $16 < \nu_T < 32$ GeV, and for charged hadrons with $0.7 < z_T < 0.8$, where $\nu_T$ and $z_T$ are the 'true' $\nu, z$ values. $\nu_s$ and $z_s$ denote the 'observed' $\nu, z$ values.

| $\nu_s$ [GeV] | $\overline{\nu}$Ne | | | | |
|---|---|---|---|---|---|
| | $z_s$ | | | | |
| | $0.5 - 0.6$ | $0.6 - 0.7$ | $0.7 - 0.8$ | $0.8 - 0.9$ | $0.9 - 1.0$ |
| $8 - 16$ | 0.009 | 0.009 | 0.009 | 0.016 | 0.000 |
| $16 - 32$ | 0.136 | 0.297 | 0.257 | 0.055 | 0.014 |
| $32 - 64$ | 0.055 | 0.024 | 0.016 | 0.003 | 0.000 |
| $\nu_s$ [GeV] | $\overline{\nu}$p | | | | |
| | $z_s$ | | | | |
| | $0.5 - 0.6$ | $0.6 - 0.7$ | $0.7 - 0.8$ | $0.8 - 0.9$ | $0.9 - 1.0$ |
| $8 - 16$ | 0.000 | 0.000 | 0.010 | 0.031 | 0.003 |
| $16 - 32$ | 0.068 | 0.313 | 0.343 | 0.078 | 0.010 |
| $32 - 64$ | 0.041 | 0.007 | 0.000 | 0.000 | 0.000 |

# Figure Captions

Fig. 1: Fragmentation functions $D(z)$ of charged hadrons for the raw data in (anti)neutrino Ne (solid circles) and (anti)neutrino proton (open circles) interactions for four intervals of $\nu$, using method [19]. The dotted lines are the predictions for the proton target from the Lund model [6] using $C = 1.8$ in eq. (9). a) neutrino, b) antineutrino interactions.

Fig. 2: Corrected $R_{\mathrm{Ne}}(z)$ for charged hadrons, in the kinematic region $4 < \nu < 8$ GeV. a) neutrino, b) antineutrino interactions. The crosses on the error bars correspond to the unfolded $R_{\mathrm{Ne}}$ from the two energy estimation methods [18] and [19]. The lines represent the predictions of the $l_c$ (solid and dash-dotted lines) and the $l_y$ (dashed and dotted lines) model. The dotted and the dash-dotted curves are for $\sigma_q = 0$. The full and the dashed lines represent $\sigma_q = 3$mb.

Fig. 3: Corrected $R_{\mathrm{Ne}}(z)$ for charged hadrons, in the kinematic region $8 < \nu < 16$ GeV. a) neutrino, b) antineutrino interactions. The crosses on the data points and the lines are as in Fig. 2.

Fig. 4: Corrected $R_{\mathrm{Ne}}(z)$ for charged hadrons, in the kinematic region $16 < \nu < 32$ GeV. a) neutrino, b) antineutrino interactions. The crosses on the data points and the lines are as in Fig. 2.

Fig. 5: Corrected $R_{\mathrm{Ne}}(z)$ for charged hadrons, in the kinematic region $32 < \nu < 64$ GeV for neutrino interactions. The crosses on the data points and the lines are as in Fig. 2.

Fig. 6: Corrected $R_{\mathrm{Ne}}(z > 0.2) = (\int_{0.2}^{1} D_{\mathrm{Ne}}(z)dz)/(\int_{0.2}^{1} D_{\mathrm{p}}(z)dz)$ for charged hadrons as a function of $\nu$ in the range $4 < \nu < 64$ GeV. In all $\nu$ bins neutrino and antineutrino data have been combined. The crosses on the error bars correspond to the unfolded $R_{\mathrm{Ne}}$ from the two energy estimation methods [18] and [19]. The dotted and the dash-dotted curves are for $\sigma_q = 0$, the full and the dashed ones for $\sigma_q = 3$ mb.

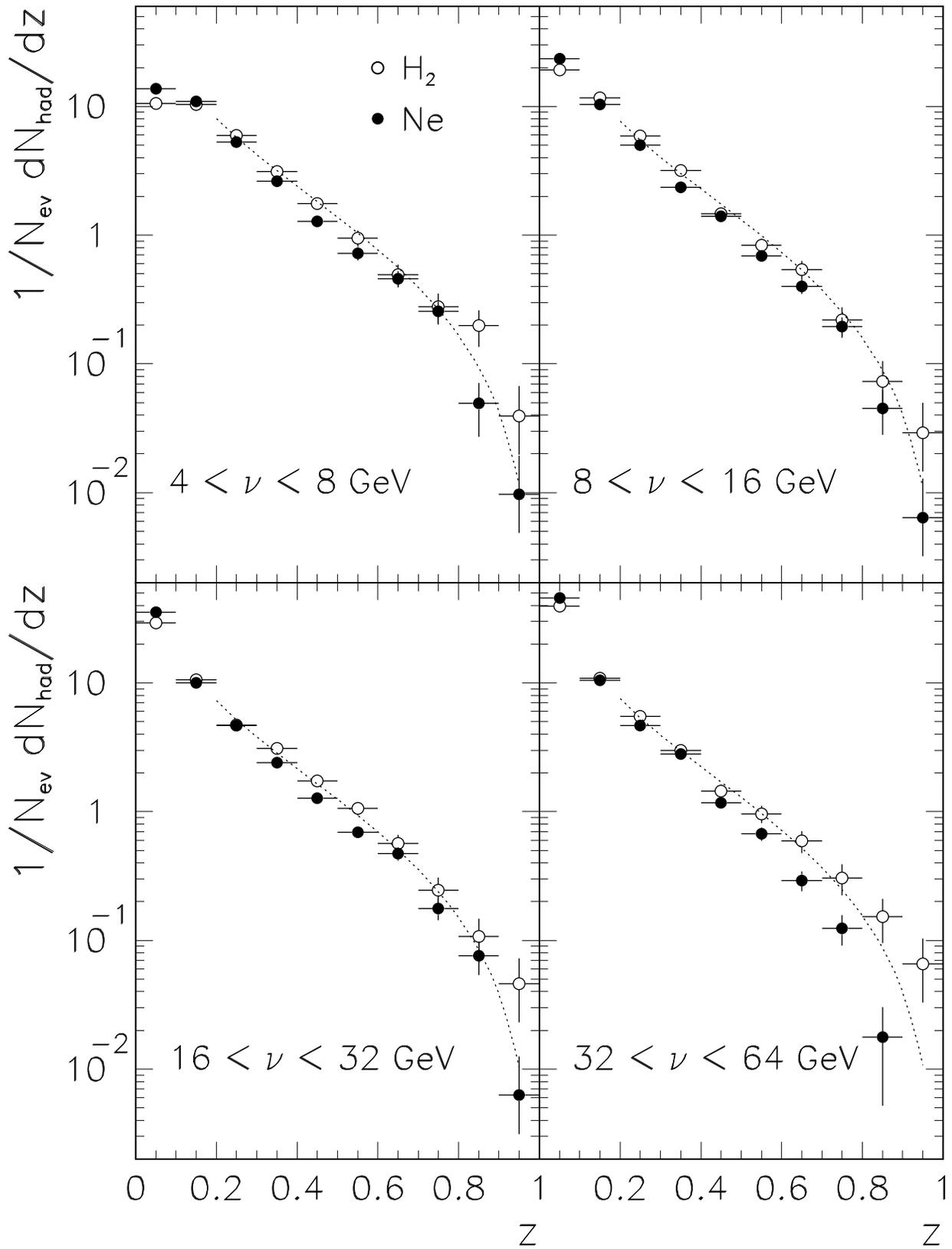

Figure 1a

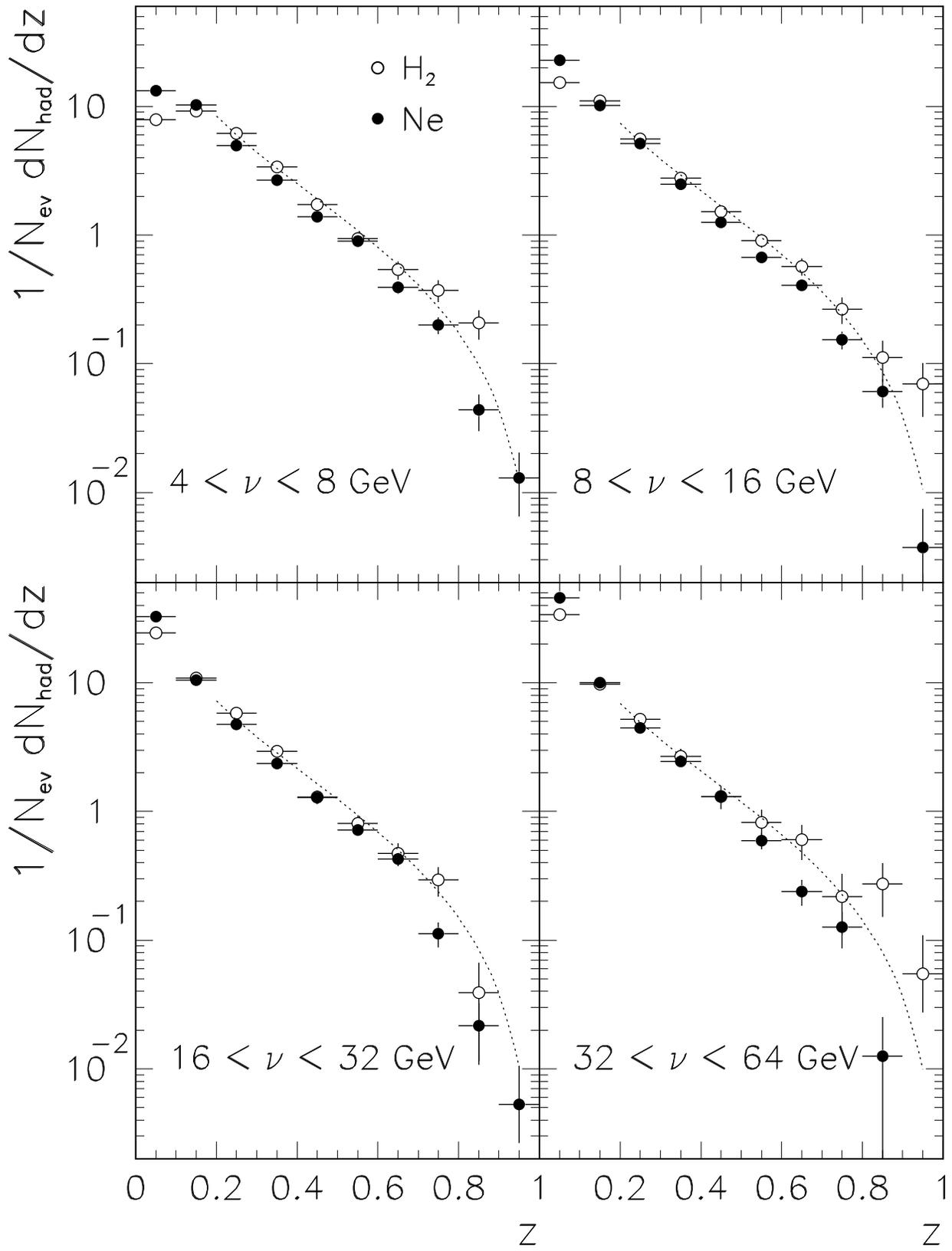

Figure 1b

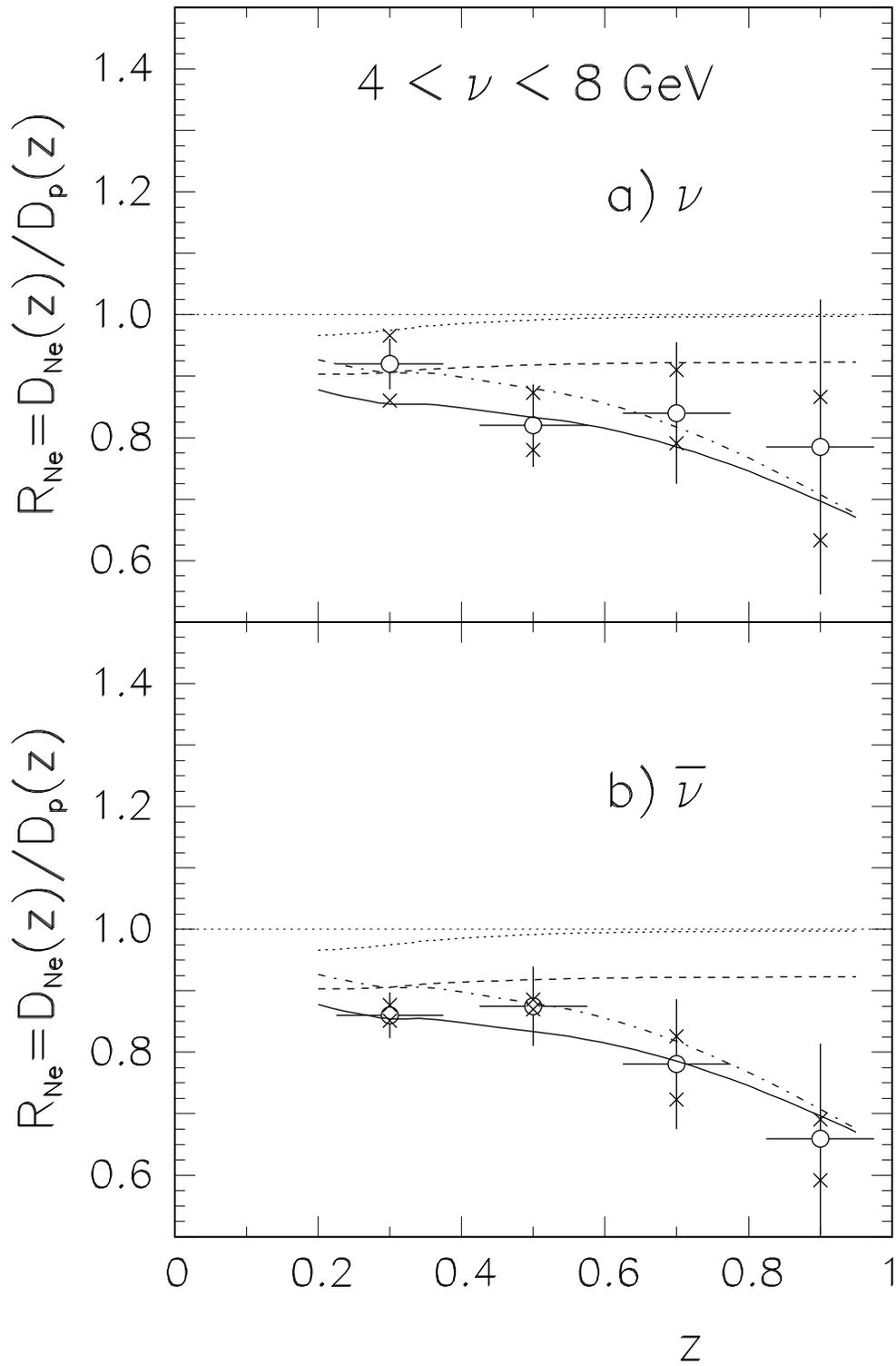

Figure 2

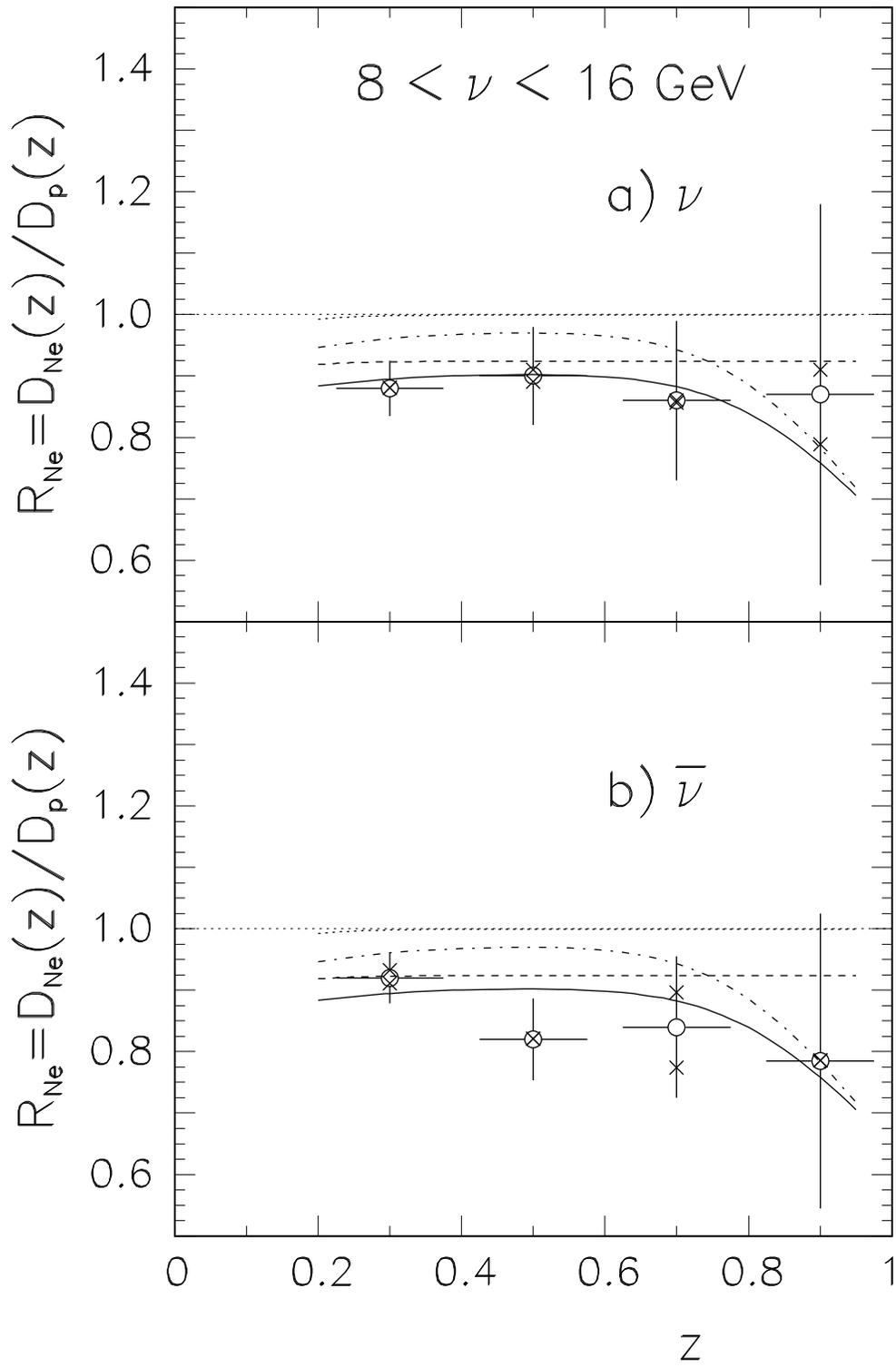

Figure 3

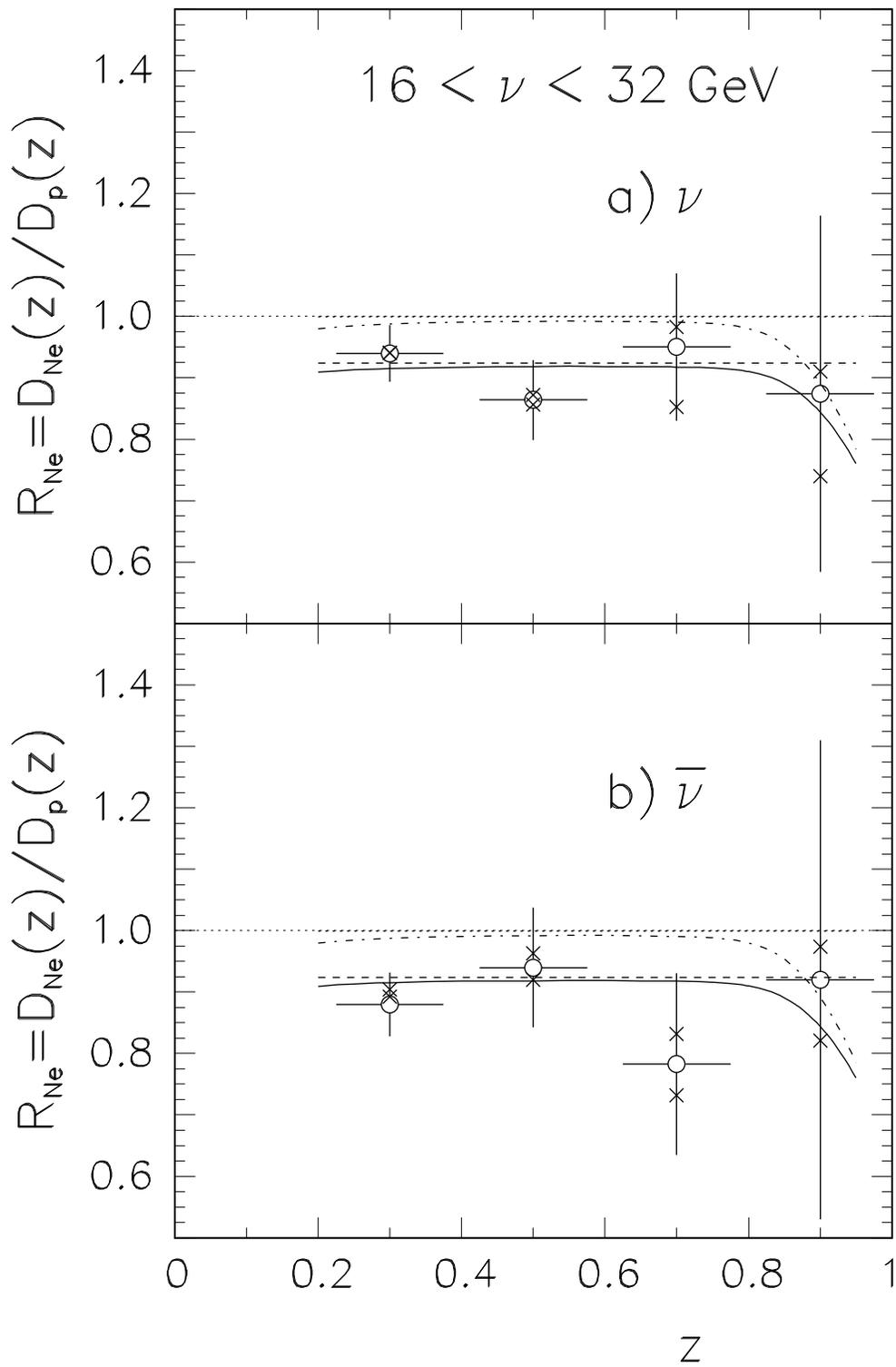

Figure 4

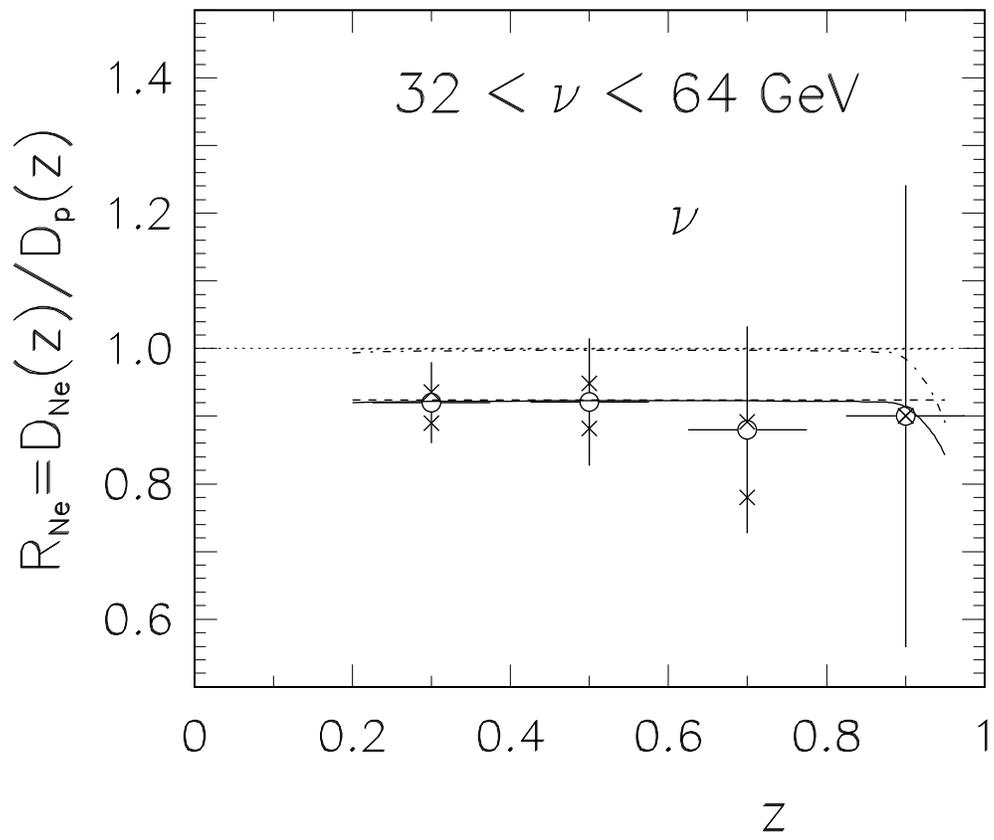

Figure 5

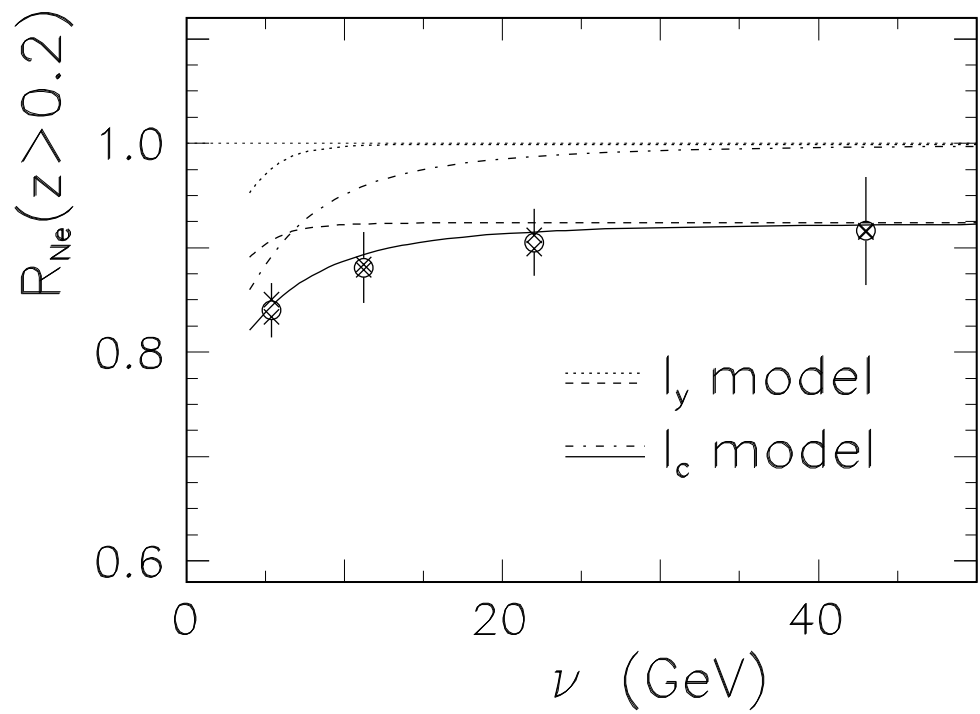

Fig. 6